\documentclass[aps,pre,twocolumn, showpacs,superscriptaddress,floatfix]{revtex4}
\usepackage{graphicx}
\usepackage{amsmath}
\usepackage{amsthm}
\usepackage{multirow}
\usepackage{relsize}
\usepackage{amssymb}
\usepackage{bm}
\usepackage{textcomp}

\usepackage{epstopdf}

\DeclareMathAlphabet{\mathitbf}{T1}{cmr}{bx}{it} 
\begin{document}

\title{Universal behavior of crystalline membranes: crumpling transition and Poisson ratio of the flat phase}

\author{R.~Cuerno} \affiliation{Departamento de Matem\'aticas and Grupo
  Interdisciplinar de Sistemas Complejos (GISC), Universidad Carlos III, 28911
  Legan\'es, Spain} \author{R.~Gallardo Caballero} \affiliation{Departamento
  de Electr\'onica and Instituto de Computaci\'on Cient\'{\i}fica Avanzada
  (ICCAEx), Universidad de Extremadura, 06071 Badajoz, Spain.}
\author{A.~Gordillo-Guerrero} \affiliation{Departamento de Electr\'onica and
  Instituto de Computaci\'on Cient\'{\i}fica Avanzada (ICCAEx), Universidad de
  Extremadura, 06071 Badajoz, Spain.}  \affiliation{Instituto de
  Biocomputaci\'on y F\'{\i}sica de Sistemas Complejos (BIFI), 50018 Zaragoza,
  Spain.}  \author{P.~Monroy} \affiliation{Instituto de Física
  Interdisciplinar y Sistemas Complejos IFISC (CSIC-UIB) Campus Universitat de
  les Illes Balears E-07122 Palma de Mallorca. Spain}
\author{J.~J.~Ruiz-Lorenzo} \affiliation{Departamento de F\'{\i}sica and
  Instituto de Computaci\'on Cient\'{\i}fica Avanzada (ICCAEx), Universidad de
  Extremadura, 06071 Badajoz, Spain.}  \affiliation{Instituto de
  Biocomputaci\'on y F\'{\i}sica de Sistemas Complejos (BIFI), 50018 Zaragoza,
  Spain.}

\date{\today}

\begin{abstract}
We revisit the universal behavior of crystalline membranes at and below the
crumpling transition, which pertains to the mechanical properties of important
soft and hard matter materials, such as the cytoskeleton of red blood cells or
graphene. Specifically, we perform large-scale Monte Carlo simulations of a
triangulated two-dimensional phantom network which is freely fluctuating in
three-dimensional space. We obtain a continuous crumpling transition
characterized by critical exponents which we estimate accurately through the
use of finite-size techniques. By controlling the scaling corrections, we
additionally compute with high accuracy the asymptotic value of the Poisson
ratio in the flat phase, thus characterizing the auxetic properties of this
class of systems.  We obtain agreement with the value which is universally
expected for polymerized membranes with a fixed connectivity.
\end{abstract}

\pacs{64.60.-i,64.60.F, 68.35.Rh, 64.60.-i, 87.15.Zg}

\maketitle

\section{Introduction}

Crystalline or polymerized membranes (CM) constitute a natural two-dimensional
generalization of the simple idea behind one-dimensional polymeric
chains. Namely, a two-dimensional arrangement of monomers, which are connected
by rigid bonds that never break \cite{Bowick01}.  CM are expected to provide
good approximations to the mechanical properties of a number of interesting
two-dimensional material systems.  Among others \cite{Bowick96}, they account
for the thermal fluctuations of the cytoskeleton of red blood cells
\cite{Schmidt93,Boal12} and provide an accurate first step to describe the
unique mechanical properties of graphene
\cite{Katsnelson12,Katsnelson2013}. Such mechanical features include e.g.\ the
existence of intrinsic, thermally-induced ripples in the graphene sheet
\cite{Fasolino07,Braghin10}, which can strongly influence the electronic and
magnetic properties \cite{Katsnelson12} of this important two-dimensional
material.

In thermal equilibrium, the phase diagram of crystalline membranes possesses a
number of remarkable properties \cite{Nelsonetal04}. Thus, in contrast with
polymers in solution \cite{DesCloizeaux90}, CM are flat (albeit rough) at low
temperatures, namely, the local normal directions at different points of the
membrane have a well-defined average orientation. Nevertheless, there is a
phase transition to a {\em crumpled} morphology at a finite value of
temperature, above which the local normal directions are disordered. For the
case of {\em phantom} CM, namely, in the absence of self-avoidance, both,
theoretical and numerical studies confirm this behavior
\cite{Bowick01,Nelson04,Gompper04}, which also differs from what is found for
{\em fluid} membranes \cite{Leibler04}. For self-avoiding membranes in
physical dimensions, namely, a two-dimensional network fluctuating in
three-dimensional space, there is no crumpling transition
\cite{Bowick01_b}. Rather, a unique phase exists for all temperatures
which is flat, with similar characteristics to the low-temperature phase
of phantom membranes.

The flat phase of crystalline membranes also hosts another remarkable property
\cite{Falcioni97}, namely, {\em auxetic} behavior, which is signaled by a
negative Poisson ratio $\sigma$ \cite{Greaves11}. This property implies that
the membrane expands transversely when stretched longitudinally, contrary with
experience with most common elastic materials. Auxetic materials are expected
to have good mechanical properties \cite{Greaves11}, such as high energy
absorption and fracture resistance; see in particular \cite{Grima15} for a
recent study of the potential of graphene from this point of view, as assessed
by Molecular Dynamics simulations.  Moreover, the conjectured negative value
of $\sigma$ seems to be independent of the occurrence of self-avoidance
constraints, $\sigma=-1/3$ ---as obtained within the so-called self-consistent
screening approximation (SCSA) \cite{Doussal92}--- having been put as
characteristic of a unique universality class for fixed-connectivity membranes
\cite{Bowick01_c}.

In spite of these interesting properties of the flat phase of crystalline
membranes, they remain to be fully understood. For instance, the nature of the
crumpling transition is still a subject of debate. While the computational and
analytical works quoted above mostly suggested that it is a continuous
transition, more recent results suggest, rather, a first order
transition. Numerical results supporting the latter conclusion have been
obtained for a number of CM-related models in e.g.\ Refs.\ \cite{Kownacki02}
(with a truncated Lennard-Jones potential) and \cite{Koibuchi04,Koibuchi05,Koibuchi14}
(spherical topology). Results from non-perturbative renormalization group (RG)
calculations are in agreement with these \cite{Essafi14}. Nevertheless,
analogous RG studies \cite{Kownacki09,Braghin10,Hasselmann11} seem to still
favor a continuous, or perhaps weak first order, crumpling transition.

In this paper we revisit Monte Carlo simulations of a discrete model of
two-dimensional phantom membranes fluctuating in three dimensions
\cite{Bowick96}. By implementing our simulations in graphics processor units
(GPU), we are able to reach large system sizes; we further implement an
enhanced statistical analysis of data. Our results favor the view of the
crumpling transition as a continuous one and feature clear-cut convergence to
the expected $-1/3$ value of the Poisson ratio characterizing the universal
auxetic properties of the flat phase \cite{Doussal92}.

The paper is organized as follows. In Sec.\ II we recall the basics of the
continuum description of phantom membranes, together with a detailed
connection to the discrete model to be simulated numerically. Section III
details the observables to be evaluated and their finite-size analysis. Our
simulation and statistical analysis methods are described in Sec.\ IV, after
which numerical results are presented in Sec.\ V. We discuss our main results
and summarize our conclusions in Sec.\ VI. Three appendices are provided for
details on some elasticity equations, renormalization group (RG) estimates for
the correction-to-scaling exponent, and on the practical implementation of our
simulation code in GPUs.

\section{Model}

\subsection{Continuum description}

For the sake of later reference, we briefly recall here the basics of the
Landau description of (phantom) polymerized membranes in physical
dimensions. Thus, a two-dimensional membrane fluctuating in three-dimensional
space can be described geometrically as a vector field $\vec{r}(x^1,x^2) \in
\mathbb{R}^3$, with $\mathitbf{x}=(x^1, x^2) \in \mathbb{R}^2$. The tangent
vectors are $\vec{t}_\alpha = \partial_\alpha \vec{r}$, where, as for all
additional Greek indices in this work, $\alpha=1,2$. These vectors allow us to
define the metric tensor in the usual way \cite{David04},
\begin{equation}
g_{\alpha\beta}\equiv
\partial_{\alpha}\vec{r}\cdot\partial_{\beta}\vec{r}\, ,
\end{equation}
so that
$d^2 s=d^2 x \sqrt{g}$.

The most general form of the Landau free energy can be written using general
principles \cite{Paczuski88,Nelson04}, namely: locality and translational
invariance (implying dependence only on the local values of the tangent
vectors $\vec{t}_\alpha$ and their derivatives), rotational symmetry in
$\mathbb{R}^3$ (implying dependence on scalar products of the tangent vectors
and their derivatives, and on even powers of these) and translational and
rotational invariance in $\mathbb{R}^2$. Thus, one can write
\cite{Paczuski88,Nelson04}
\begin{eqnarray}\label{ELandau}
\nonumber
    F[\vec{t}_{\alpha}(\mathitbf{x}),T]&=&
\int d^2 s
    \left[
      \frac{t}{2}(\vec{t}_{\alpha})^2+
      u(\vec{t}_{\alpha}\vec{t}_{\beta})^2+
      v(\vec{t}_{\alpha}\vec{t}^{\alpha})^2 \right.\\
 &+&
     \left. \frac{\kappa}{2}(\partial_{\alpha}\vec{t}_{\alpha})^2 \right] ,
\end{eqnarray}
where self-avoidance is neglected \cite{Bowick01}.

One can parameterize the three-dimensional coordinates $\vec{r}$ of points on
a {\em flat membrane} taking as a reference the base plane, namely, the plane
determined by the average position of all the atoms. Denoting the equilibrium
position on this plane by $\mathitbf{x}$, and in-plane and perpendicular
fluctuations by $\mathitbf{u}$ and $h$, respectively, we have
\begin{equation}
\vec{r}(\mathitbf{x})=(\mathitbf{x}+\mathitbf{u(x)},h(\mathitbf{x})),
\end{equation}
where $\langle (\mathitbf{u(\mathitbf{x})},h(\mathitbf{x}))\rangle =0$.

Moreover, on a flat membrane the strain tensor can be written as \cite{Nelson04}
\begin{equation}
u_{\alpha\beta}=\frac{1}{2}\left(\partial_{\alpha}\vec{r}\cdot\partial_{\beta}\vec{r}
-\delta_{\alpha\beta}\right)
=\frac{1}{2}\left(g_{\alpha \beta}-\delta_{\alpha\beta}\right)\,.
\end{equation}
Taking into account the in-plane ($\mathitbf{u}$) and out-plane ($h$)
displacements, and neglecting quadratic terms in $u_{\alpha}$, we can write
\begin{equation}
u_{\alpha\beta}=\frac{1}{2}\left( \partial_\alpha u_\beta + \partial_\beta
u_\alpha +\partial_\alpha h \partial_\beta h \right) .
\end{equation}
The so-called elastic part of the Landau free energy \eqref{ELandau}, namely,
\begin{equation}
    F_E[\vec{t}_{\alpha}(\mathitbf{x}),T]=
\int d^2 s
    \left[
      \frac{t}{2}(\vec{t}_{\alpha})^2+
      u(\vec{t}_{\alpha}\vec{t}_{\beta})^2+
      v(\vec{t}_{\alpha}\vec{t}^{\alpha})^2 \right] ,
\end {equation}
can now be written as
\begin{equation}
\label{eq:elastic}
F_E(\vec{r})=\int d^2 x f_E(\vec{r})=
\int d^2 x
\left[
\mu u_{\alpha\beta}u^{\alpha\beta} +
\frac{\lambda}{2}(u_{\alpha}^{\alpha})^2\right]\, ,
\end{equation}
where $\lambda$ and $\mu$ are the Lam\'e coefficients \cite{metric}.
The remaining part of the Landau free energy, namely,
\begin{equation}
    F_C[\vec{t}_{\alpha}(\mathitbf{x}),T]=
\int d^2 s
     \left[ \frac{\kappa}{2}(\partial_{\alpha}\vec{t}_{\alpha})^2 \right]
\end{equation}
is a curvature contribution that can be written as
\begin{equation}
  F_C[\vec{t}_{\alpha}(\mathitbf{x}),T]=2 \kappa
\int d^2 x K_\alpha ^\beta K ^\alpha _\beta ,
\end{equation}
where indeed $K_\alpha ^\beta$ is the extrinsic curvature \cite{extrinsic}.

\subsection{Discrete model}

\begin{figure}[t!]
\centering
\includegraphics[width=0.80\columnwidth, angle=0]{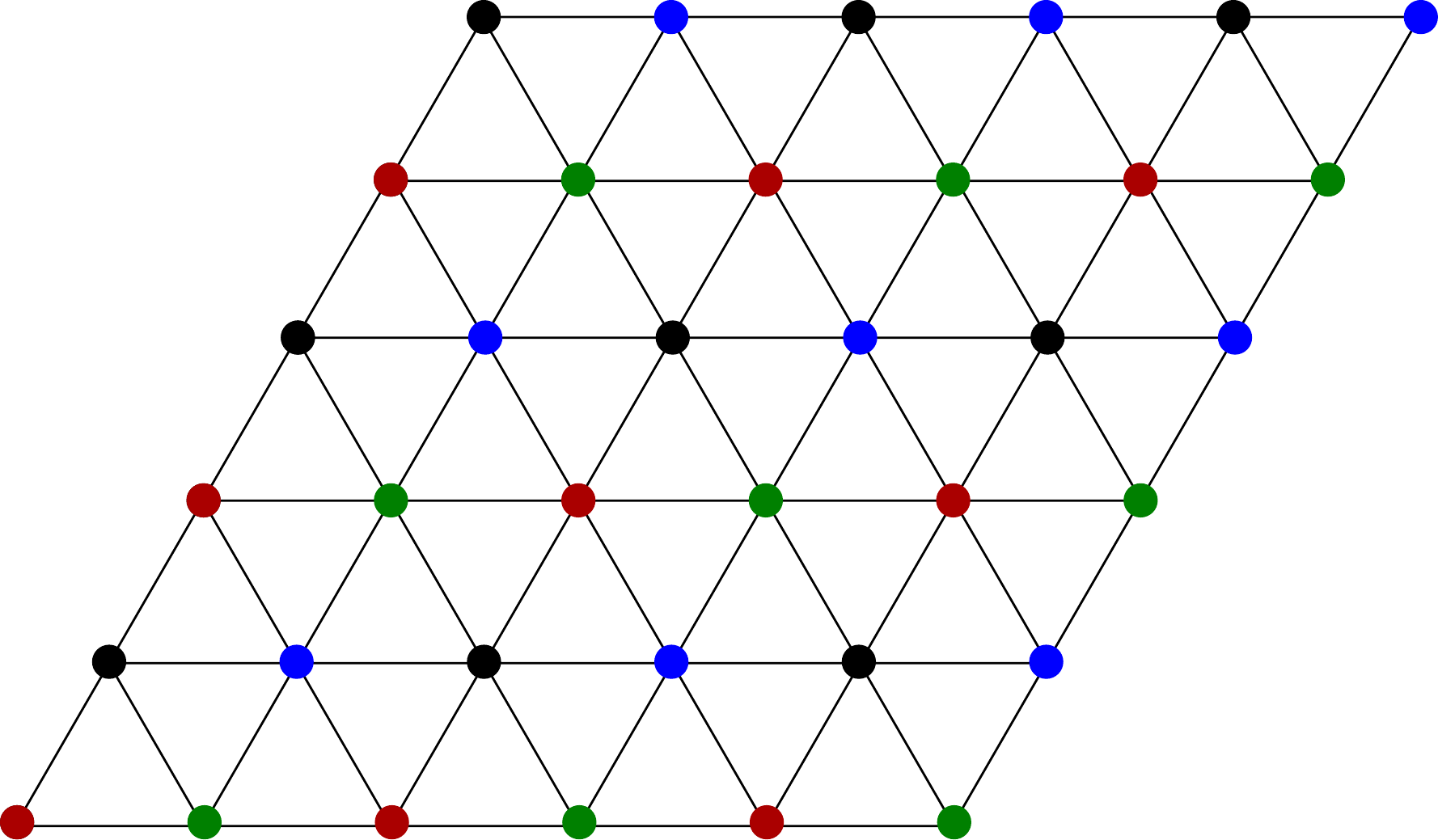}
\caption{(color online) Lattice that has been employed in this work, for
  $L=6$, in which a four-color partition is illustrated. Points with the same
  color can be updated simultaneously. This partition has been used to speed
  up, in a highly parallel way, our numerical simulations using GPU
  processors.}
\label{fig:GPU}
\end{figure}

\begin{figure}[h!]
\centering
\includegraphics[width=0.80\columnwidth, angle=0]{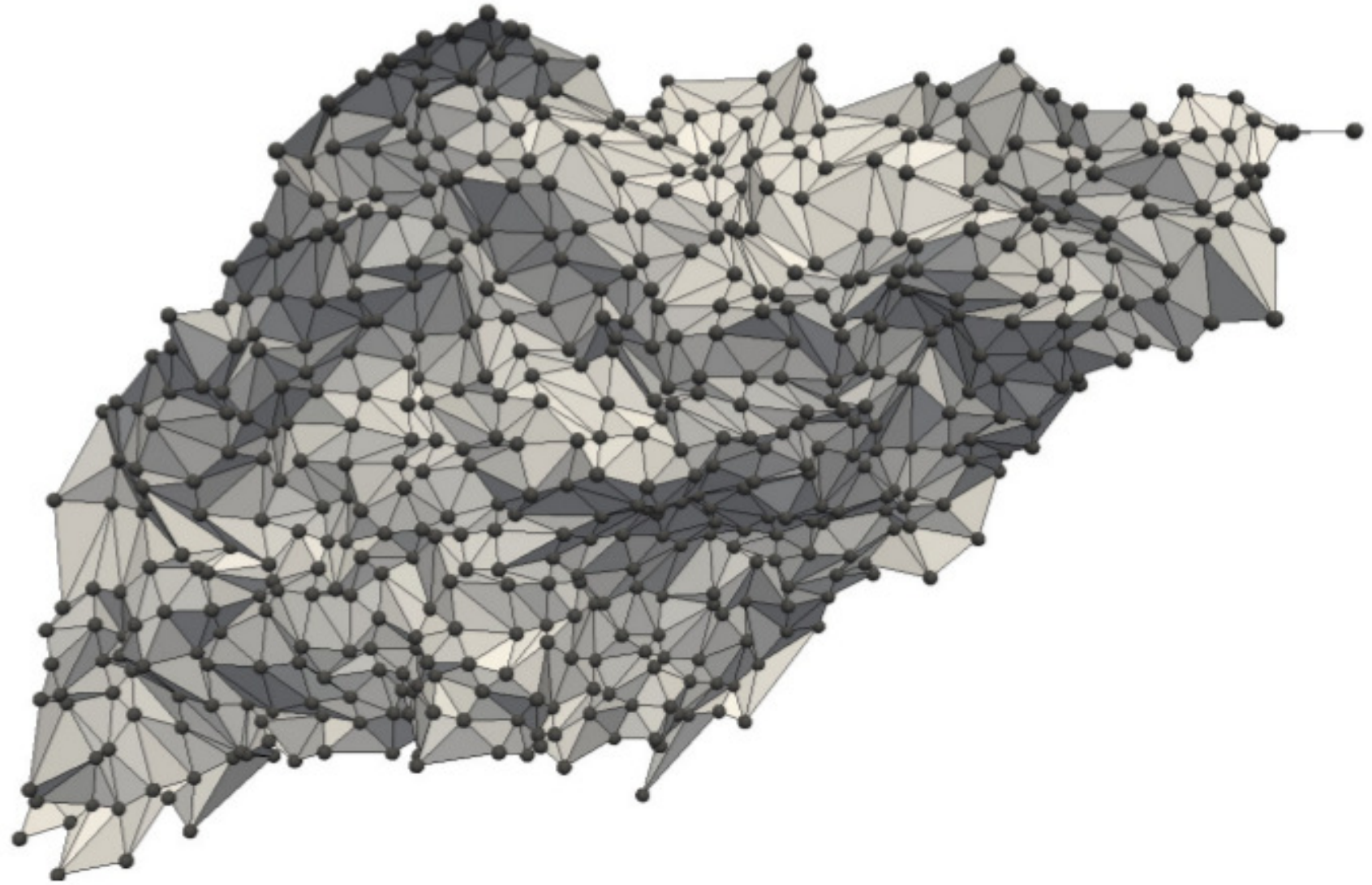}
\caption{(color online) Sample membrane configuration at $\kappa=2$ in the flat phase, for $L=24$.}
\label{fig:Snapshot}
\end{figure}

We define our computational model on an $N=L\times L$ two-dimensional
triangular lattice like the one shown in see Fig.\ \ref{fig:GPU}, embedded in
three-dimensional space, see Fig.\ \ref{fig:Snapshot}. The position of the
points is thus labeled by a three-dimensional vector $\vec{r}$. We also define
triangular plaquettes with associated normal vectors denoted as $\vec{n}$. In
general, we will denote points using the indices $i$, $j$, $k$, etc.\ and
plaquettes using $a$, $b$, $c$, etc.

The Hamiltonian we study is
\begin{equation}
{\cal H}= {\cal H}_E+ \kappa {\cal H}_C ,
\end{equation}
which includes an elastic part, ${\cal H}_E$, given by
\begin{equation}
{\cal H}_E= \frac{1}{2} \sum_{<ij>} |\vec{r}_i-\vec{r}_j|^2 ,
\end{equation}
and a curvature  contribution, ${\cal H}_C$, which reads
\begin{equation}
{\cal H}_C= \frac{1}{2} \sum_{<ab>} |\vec{n}_a-\vec{n}_b|^2 .
\end{equation}
As usual, $<ij>$ denotes nearest-neighbor points, while $<ab>$ denotes
nearest-neighbor plaquettes.

Within a mean-field approach, it is possible to show
\cite{Seung88,Bowick96,Monroy13} that the above Hamiltonians retrieve the
corresponding contributions to the Landau free energy in the continuum limit,
namely,
\begin{equation}
{\cal H}_E \to \int d^2 s \left( \frac{1}{2} u_\alpha ^\alpha u_\beta^ \beta +
u_{\alpha \beta} u^{\alpha \beta}  \right) ,
\end{equation}
\begin{equation}
{\cal H}_C \to \int d^2 s K_\alpha ^\beta K ^\alpha _\beta .
\end{equation}

\section{Observables and the Finite Size Scaling Method}

The specific heat can be computed as \cite{Harnish1991}
\begin{equation}
C_V=\frac{\kappa^2}{N} \left( \langle E_C^2 \rangle -
\langle E_C\rangle^2  \right) \,,
\end{equation}
where
\begin{equation}
E_C=\sum_{<ab>} \vec{n}_a \cdot \vec{n}_b \,,
\end{equation}
is the curvature energy. The behavior of the specific heat near the crumpling
transition that takes place at $\kappa=\kappa_c$ is characterized as,
\begin{equation}
C_V\sim |\kappa-\kappa_c|^{-\alpha}\,.
\end{equation}

The correlation length is defined using the correlation among the normals of
the system,
\begin {equation}
\langle  \vec{n}(x)\cdot \vec{n}(0)\rangle \propto e^{-|x|/\xi}\,,
\end{equation}
the scaling law for $\xi$ being
\begin{equation}
\xi(\kappa) \sim |\kappa-\kappa_c|^{-\nu}\,.
\end{equation}
Right at the critical point, the maximum of the specific heat scales as
\begin{equation}
\label{eq:scaling_cmax}
C_{\mathrm{max}} \propto C_a+ B L^{\alpha/\nu} \,.
\end{equation}
where $C_a$ describes the contribution from the analytical part of the
free energy.

We can further describe the space configuration of the membrane by considering
the gyration radius for the distribution of surface nodes, which is defined as
\begin{equation}
R^2_g=\frac{1}{3 N} \langle \sum_i \vec{R}_i\cdot \vec{R}_i  \rangle \,,
\end{equation}
where $\vec{R}_i \equiv \vec{r}_i -\vec{r}_{\text{CM}}$, with $\vec{r}_{\text{CM}}$ the position of the center of mass of the surface. In addition, linear response theory allows us to compute its $\kappa$-derivative,
\begin{equation}
\frac{d R^2_g}{d \kappa}=\langle E_C R^2_g  \rangle - \langle
E_C \rangle \langle R^2_g \rangle \equiv \langle E_C R^2_g \rangle_c .
\end{equation}
Neglecting scaling corrections, the gyration radius scales with the system
size as
\begin{equation}
\label{eq:Rscaling}
R_g \sim L^{\nu_F} f(L^{1/\nu} (\kappa-\kappa_c)),
\end{equation}
which defines the Flory exponent $\nu_F$. This exponent is related with the
Hausdorff dimension, $d_H$, of the membrane by means of
\begin{equation}
\nu_F=\frac{2}{d_H} .
\end{equation}
Near the crumpling transition, one finds $d_H=-4/\eta$, where $\eta$ is the
anomalous dimension of the $\vec{r}$-field \cite{Espriu96}.
In the flat phase one has $d_H=2$ and $\nu_F=1$, while the high temperature rough phase features
$\nu_F=0$ and $d_H=\infty$, which is linked with a logarithmic divergence of the gyration radius, $R_G^2 \sim \log L$. Therefore, the $\kappa$-derivative of the gyration radius diverges at
the critical point $\kappa=\kappa_c$ as
\begin{equation}
\label{eq:RscalingB}
\frac{d R^2_g}{d \kappa} \propto L^{2\nu_F+1/\nu} \,,
\end{equation}
an equation that will allow us to compute numerically the Flory exponent. Note
that, using the SCSA approximation, it has been found that $\nu=0.73$ and
hence $d_H=2.74$ \cite{Doussal92}.

Finally we can estimate the Poisson ratio via (see Appendix \ref{app1})
\begin{equation}
\label{eq:PoissonCalc}
\sigma=\frac{K-\mu}{K+\mu}=-\frac{\langle g_{11} g_{22}\rangle_c}{\langle
  g_{22}^2 \rangle_c}=-\frac{\langle g_{11} g_{22}\rangle_c}{\langle
  g_{11}^2 \rangle_c}\,,
\end{equation}
where $g$ is the induced metric that can be estimated using the vector which
connects nearest neighbor points, as described in Appendix \ref{app1}. Note that
we have assumed isotropy in the last two terms in Eq.\ (\ref{eq:PoissonCalc}), specifically,
$\langle g_{11}^2 \rangle_c= \langle g_{22}^2 \rangle_c$.

\section{Numerical Simulations}

We have performed Monte Carlo (MC) simulations on triangular lattices of
different sizes using a standard Metropolis algorithm. We have performed
simulations for several months both on CPUs (Intel Core I7-3770) and GPUs
(NVIDIA Tesla C2070). While the MC sweep of each site on the lattice on CPUs
is entirely sequential, in GPUs we can parallelize the simulation using a kind
of checkerboard scheme with four colours, see Fig.\ \ref{fig:GPU} and Appendix
\ref{gpu} for details.  We have obtained a gain factor around 5x simulating on
GPUs.

We have simulated lattice sizes in the $16\leq L\leq 128$ range, using
free-boundary conditions.  The thermalization protocol has been as follow: 1)
we have always started from a flat configuration, 2) we have discarded the
first $10^6$ Metropolis sweeps, and 3) we have analyzed the remainder of the
run using a logarithmic binning check of several non local observables in the
most challenging simulations, i.e., for $\kappa \simeq \kappa_c$.

After thermalization, we have saved $10^5$ configurations with at least $10^4$
Metropolis sweeps between each pair of saved configurations. Via the
computation of the integrated autocorrelation times \cite{Amit2005} we have
checked that all our measures are fully independent. To estimate the error
bars of our observables we always used a jackknife method with 20 bins.

For each lattice size, we have considered several values of $\kappa$ around
$\kappa_c$, as well as $\kappa$ values well inside the flat phase, in general
we have simulated the range $0.75\leq \kappa \leq 3.0$.

\section{Numerical Results}

\subsection{Crumpling Transition}

\begin{figure}[t!]
\centering
\includegraphics[width=0.72\columnwidth, angle=270]{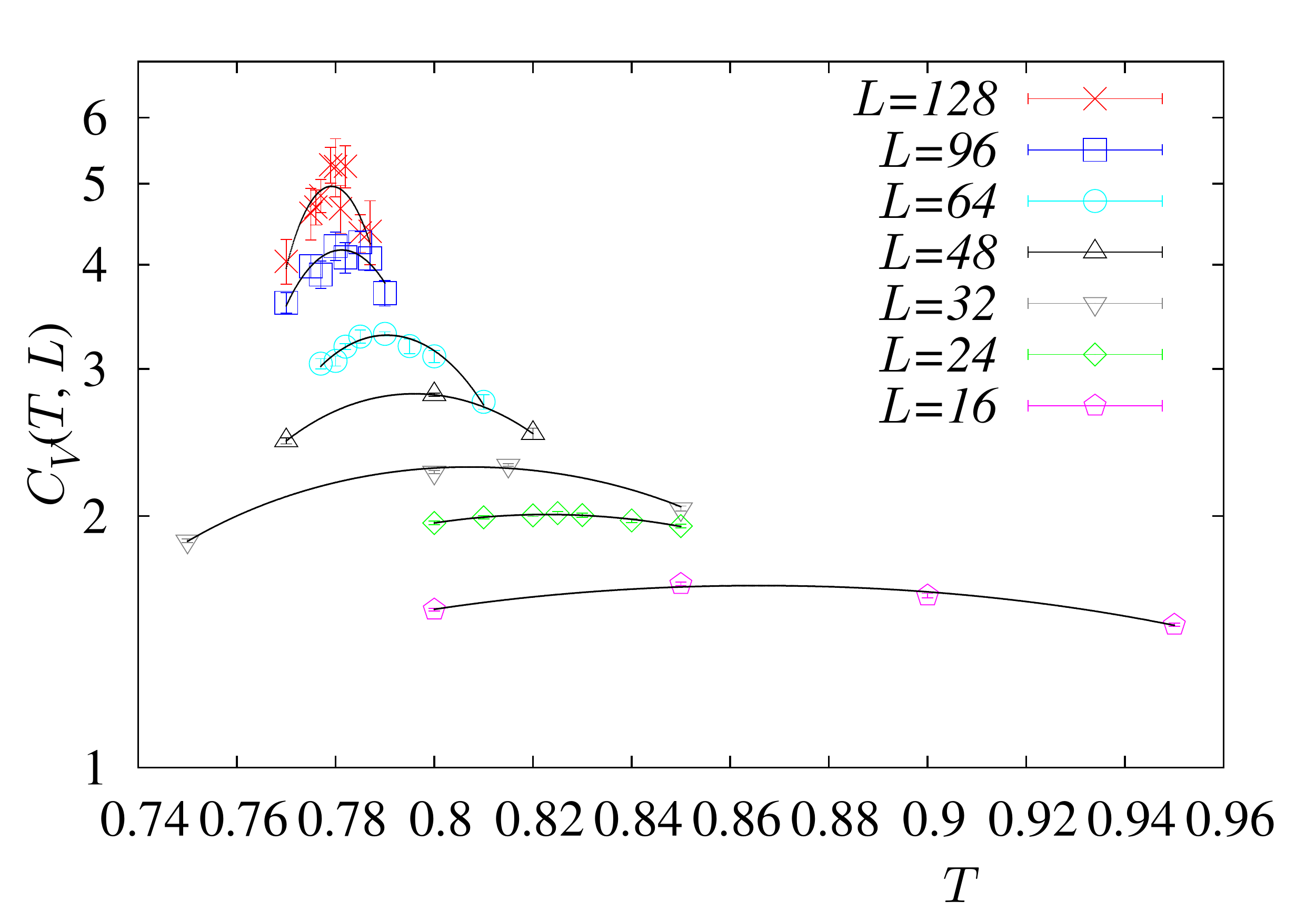}
\caption{(color online) Behavior of the specific heat as a function of
  temperature for all simulated lattice sizes. The solid black curves provide the
  least-squares fit of the points near the maximum to a quadratic form as
  described in the text. We have used  pentagons ($L=16$), rhombs ($L=24$),
  inverted triangles ($L=32$), triangles ($L=48$), circles ($L=64$), squares
  ($L=96$) and crosses ($L=128$) to mark the points in the figure.}
\label{fig:cesp}
\end{figure}

\begin{figure}[t!]
\centering
\includegraphics[width=0.72\columnwidth, angle=270]{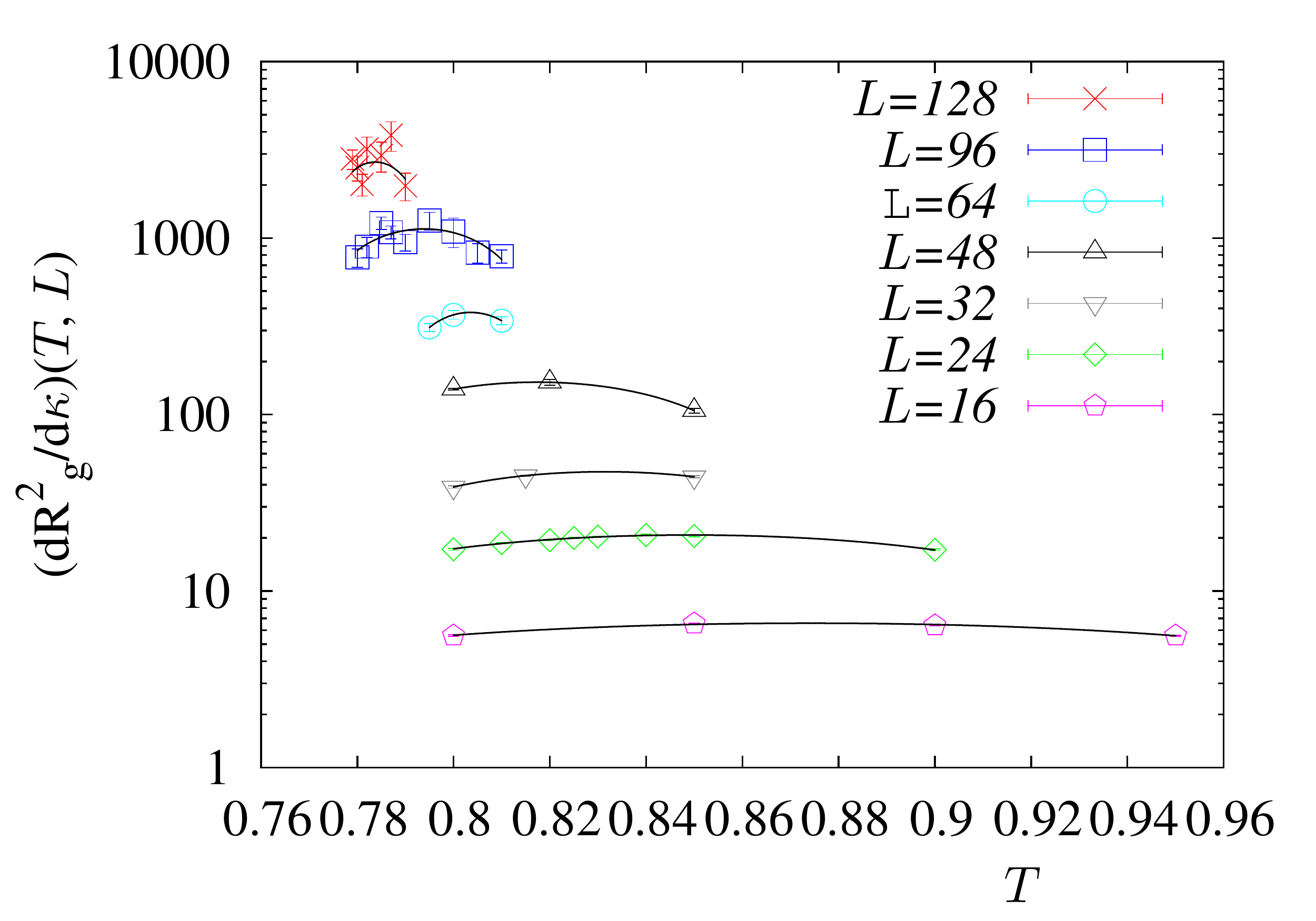}
\caption{(color online) Behavior of the $\kappa$-derivative of the squared
  gyration radius as a function of temperature for all simulated lattice
  sizes. The solid black curves provide the least-squares fit of the points near the
  maximum to a quadratic form as described in the text. We have used pentagons
  ($L=16$), rhombs ($L=24$), inverted triangles ($L=32$), triangles ($L=48$),
  circles ($L=64$), squares ($L=96$) and crosses ($L=128$) to mark the points
  in the figure.}
\label{fig:R}
\end{figure}

\begin{figure}[t!]
\centering
\includegraphics[width=0.72\columnwidth, angle=270]{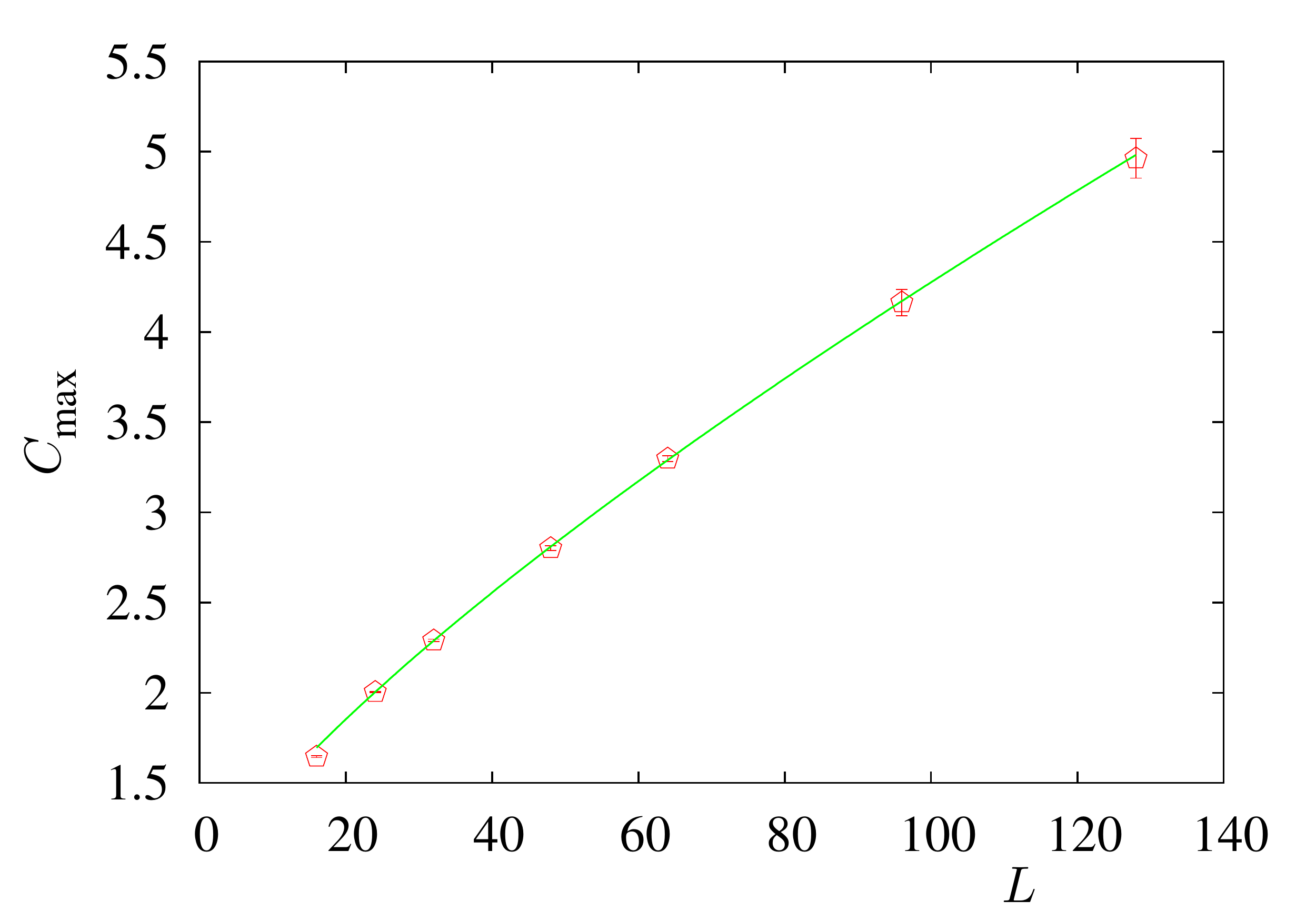}
\caption{(color online) Scaling of the maximum of the specific heat
  (pentagons) as a function of the size of the system. The solid line is a fit
  to Eq.\ (\ref{eq:scaling_cmax}).}
\label{fig:cmax}
\end{figure}

In Figs.\ \ref{fig:cesp} and \ref{fig:R} we show the behaviors of the specific
heat and $d R^2_g/d \kappa$ as functions of $\kappa$. The onset of a critical
behavior is clear in both figures. To compute the maximum of both observables,
we have fitted each curve near its maximum using least-squares (for $L<96$ we
have also used the spectral density method). Monte Carlo has been performed on
the raw data in order to compute the error bars, both in the value of the
maximum as well as in its position.

In order to characterize the critical properties (order of the phase
transition, critical exponents, etc.), we first monitor the behavior of the
maximum of the specific heat, see Fig.\ \ref{fig:cmax}. By using data with $L
> 16$, we find a clear divergence of this maximum following a power-law with a
background term, as stated in Eq.\ (\ref{eq:scaling_cmax}) above, with
$\alpha/\nu=0.756(40)$ ($\chi^2/\mathrm{d.o.f.}=0.78/3$, where d.o.f. means
the number of degree of freedom of the fit). This $\alpha/\nu=0.756(40)$ value
is definitively different from 2, characteristic of a strong first-order phase
transition, or even from 1 which would indicate a weak first-order transition
\cite{Fernandez92}. By using hyper-scaling in two dimensions [$\alpha=2 (1-
  \nu)$], we can compute simultaneously both $\alpha$ and $\nu$, to get
\begin{equation}
\label{eq:cmax-results}
\nu=0.73(1) \,\,, \, \, \alpha=0.55(2)\,.
\end{equation}

\begin{figure}[t!]
\centering
\includegraphics[width=0.72\columnwidth, angle=270]{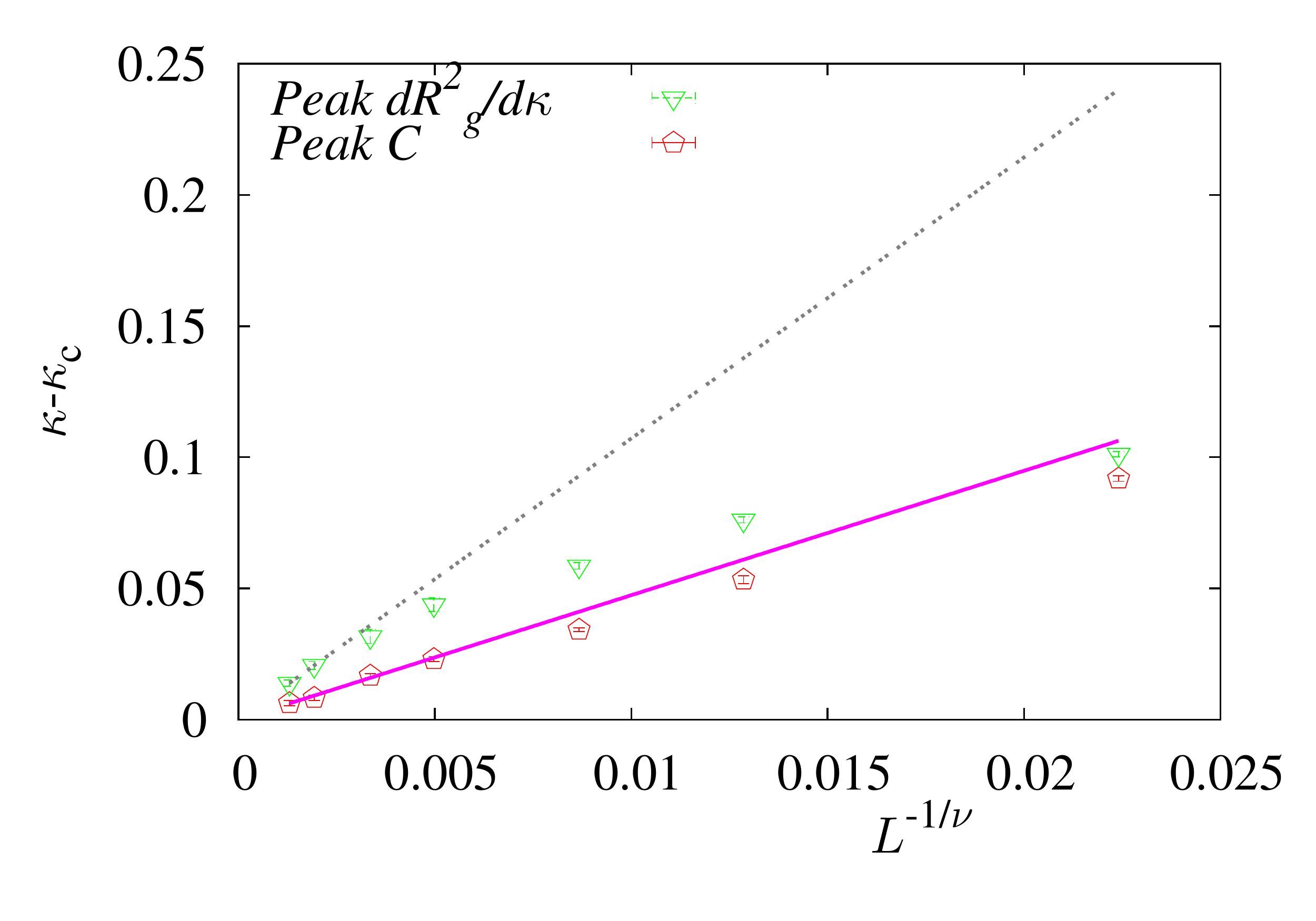}
\caption{(color online) Position of the maximum of the specific heat (lower
  data, see legend, pentagons) and of the maximum of $d R^2_g/ d\kappa$
  (inverted triangles) as a function of $L^{-1/\nu}$. In this plot we have
  employed $1/\nu=1/0.73$, as obtained from the fit of the peak of the
  specific heat. We have used the position of the specific heat maximum to
  obtain $\kappa_c$. Finally, we have fitted $d R^2_g/ d\kappa$ by imposing
  the previous obtained values of $\nu$ and $\kappa$, hence there is a single
  free parameter in the fit. The solid (specific heat) and dashed (gyraton
  radius) lines are ``linear'' fits in this scale. Notice that the gyration
  radius presents stronger scaling corrections than the specific heat.}
\label{fig:Pos_Peaks}
\end{figure}

Using the $\nu$-exponent obtained from the scaling of the maximum (see
Fig. \ref{fig:Pos_Peaks}) of the
specific heat, we can study the $\kappa$-position of this maximum, denoted as
$\kappa(L)$. This $\kappa(L)$ follows the standard scaling equation
(neglecting scaling corrections),
\begin{equation}
\kappa(L)=\kappa_c + A L^{-1/\nu} \,,
\end{equation}
where $\kappa_c$ is the infinite volume critical coupling. By using the $\nu$
value from Eq.\ (\ref{eq:cmax-results}), 
we obtain
\begin {equation}
\kappa_c=0.773(1)\,,
\end{equation}
where we have taken into account $L>32$ data only
($\chi^2/\mathrm{d.o.f.}=2.016/2$). We have been unable to compute the
correction-to-scaling exponent of the crumpling transition.

\begin{figure}[t!]
\centering
\includegraphics[width=0.72\columnwidth, angle=270]{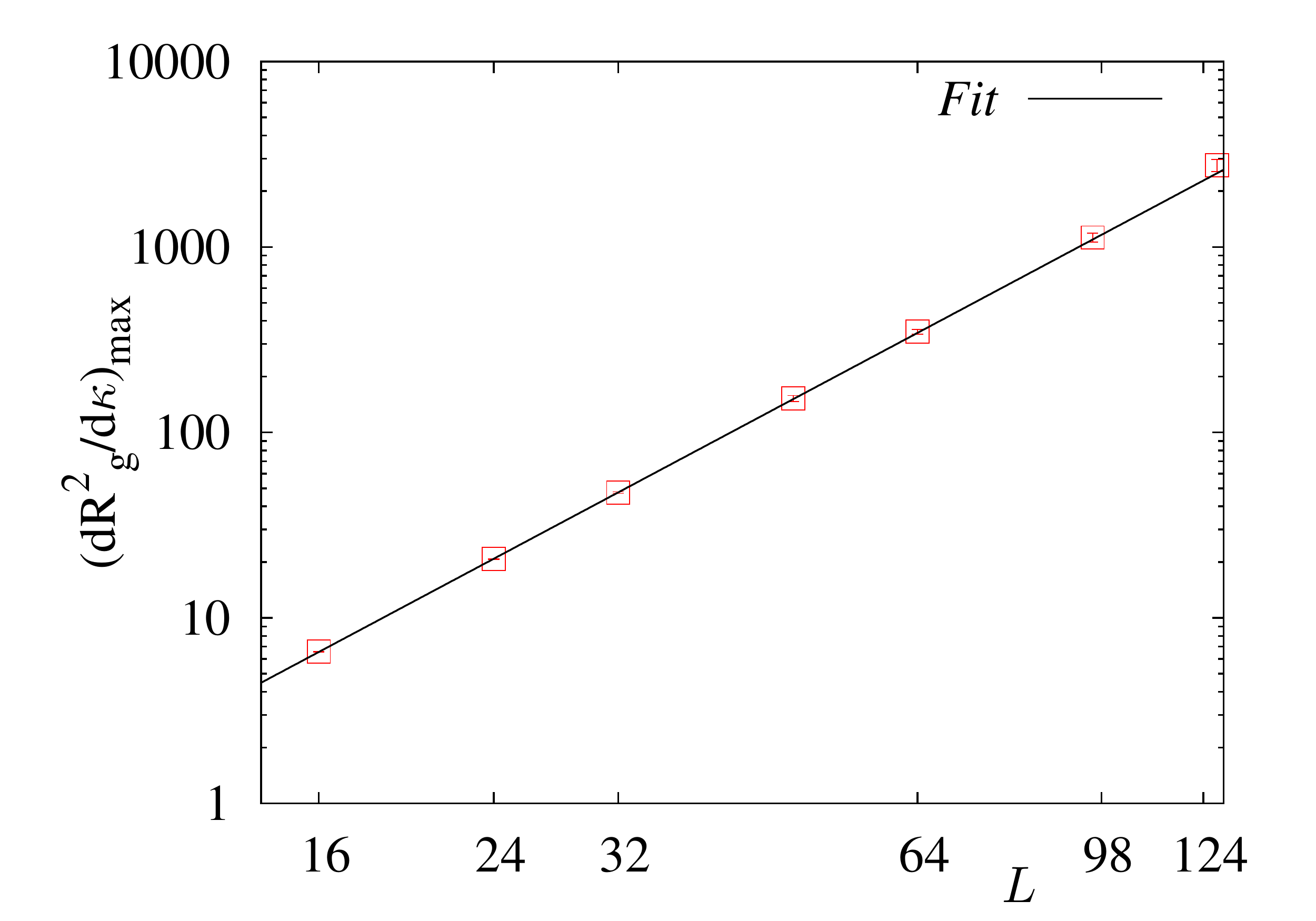}
\caption{(color online) Scaling of the maximum of the $\kappa$-derivative
  (squares) of
  the squared gyration radius, $d R^2_g/ d\kappa$, as a function of system
  size. The solid line is a fit to Eq.\ (\ref{eq:RscalingB}).}
\label{fig:DRg}
\end{figure}

We can redo the previous analysis on the derivate of the squared gyration
radius (Fig. \ref{fig:Pos_Peaks}). The
scaling of the maximum of $d R^2_g/d \kappa$ (see Fig.\ \ref{fig:DRg})
provides us with the following combination of exponents, see
Eq.(\ref{eq:RscalingB}),
\begin{equation}
2 \nu_F+\frac{1}{\nu}=2.86(1) ,
\end{equation}
with $\chi^2/\text{d.o.f.}=3.5/5$ ($L\ge 16$).
Using again the $\nu$ value obtained from the analysis of the specific heat, we can obtain the
Flory exponent, equivalently the Hausdorff dimension, of the surface at
criticality, taking into account the error bars in our value of $\nu$:
\begin{equation}
\nu_F=0.74(1) \,\,, \, \, d_H=2.70(2) \,.
\end{equation}

\subsection{Poisson ratio in the flat phase}

We next study the low-temperature (equivalently, large $\kappa$) flat phase of
phantom crystalline membranes in more detail (see Ref. \cite{Troster13} for a
detailed computation of the $\eta$-exponent). In this ordered phase,
long-range order exists in the orientation of surface normals. Moreover, this
phase is known to host auxetic behavior \cite{Bowick01}.

As mentioned in the introduction, auxetic materials have acquired a huge
importance, both from the fundamental and from the technological points of
views \cite{Greaves11}. The flat phase of CM membranes unambiguously shows a
negative value of the Poisson ratio. However, a detailed study of the scaling
corrections in this phase is still lacking, while numerical data should be
extrapolates to infinite volume in order to perform a proper comparison with
analytical results. In particular, the value $\sigma=-1/3$, obtained by the
SCSA approximation \cite{Doussal92}, has been hypothesized to characterize a
unique universality class for fixed-connectivity membranes \cite{Bowick01_c}.

\begin{figure}[t!]
\centering
\includegraphics[width=0.72\columnwidth, angle=270]{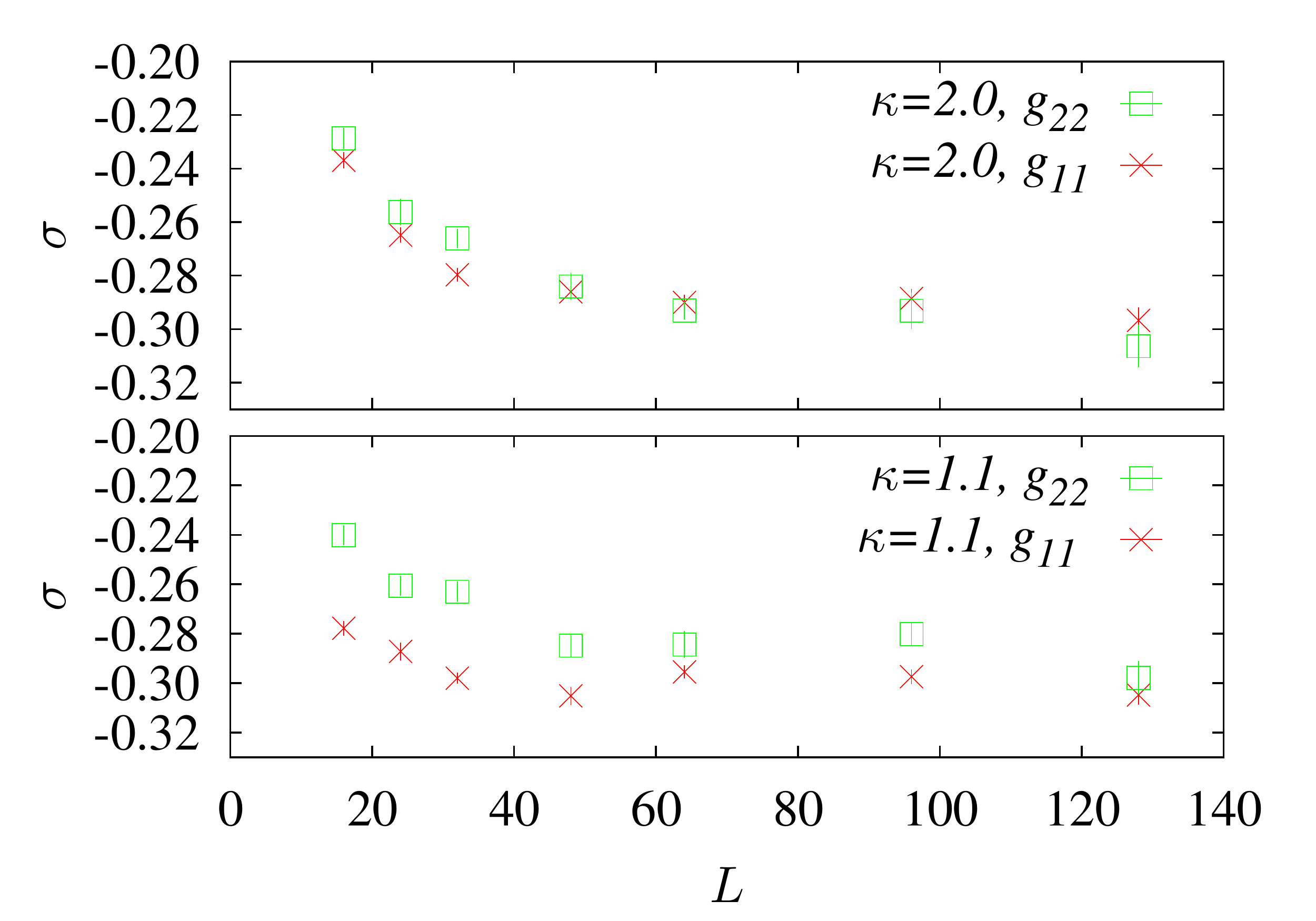}
\caption{(color online) Comparison between the computation of the Poisson
  ratio using $g_{11}$ or $g_{22}$ as the denominator in
  Eq.\ (\ref{eq:PoissonCalc}), see legend. Notice the strongly anisotropic
  behavior of the Poisson ratio for $\kappa=1.1$ (lower panel). For $\kappa=2$
  (upper panel), isotropy holds for $L > 32$.}
\label{fig:Poisson1}
\end{figure}

\begin{figure}[t!]
\centering
\includegraphics[width=0.72\columnwidth, angle=270]{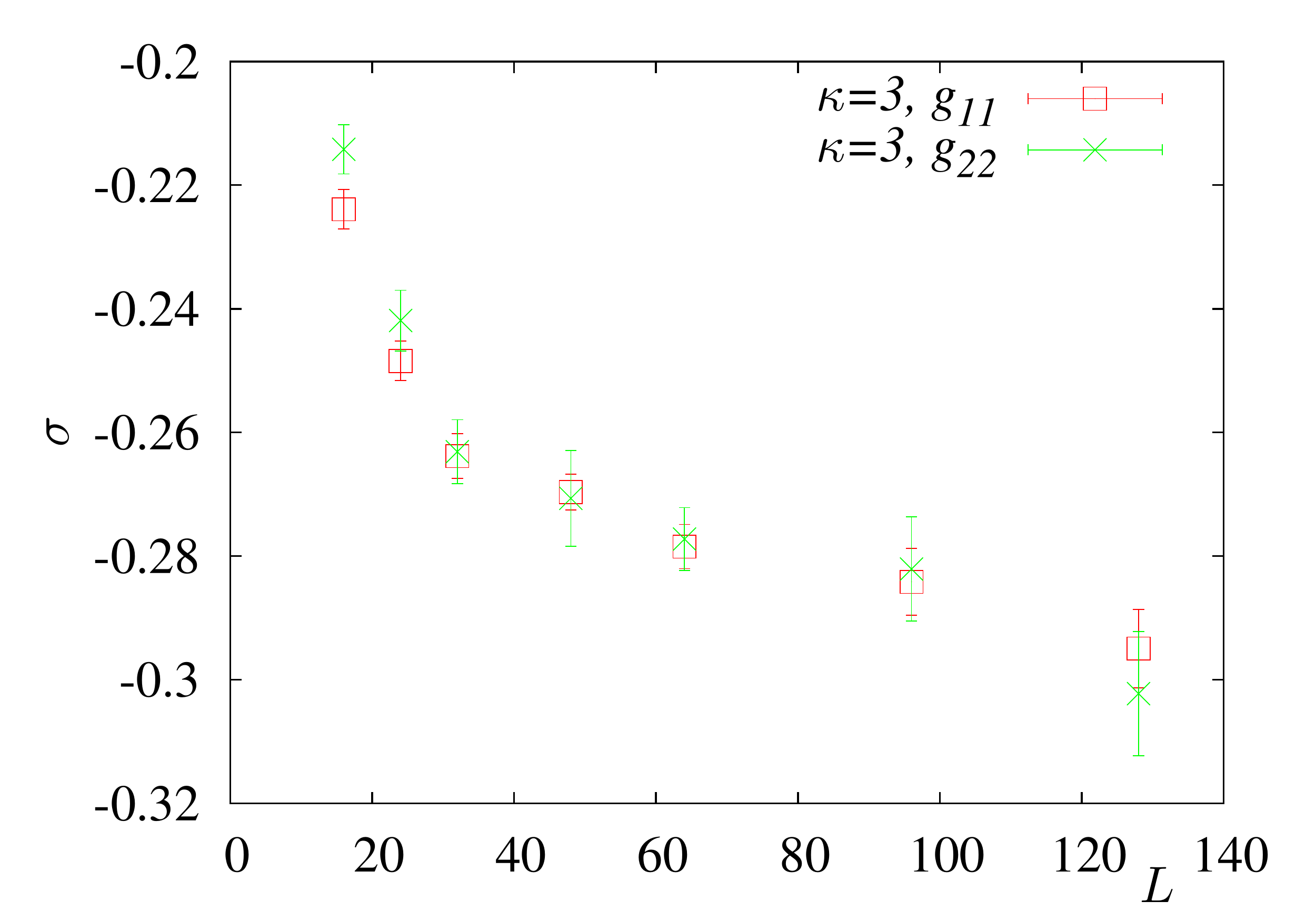}
\caption{(color online) Comparison between the computation of the Poisson
  ratio using $g_{11}$ (squares)  or $g_{22}$ (crosses) as the denominator in
  Eq.\ (\ref{eq:PoissonCalc}) for $\kappa=3$, see legend. Isotropy holds for
  $L \ge 24$.}
\label{fig:Poisson2}
\end{figure}

We have first explicitly checked the isotropy of finite-size membranes in the
flat phase. To do this, we have monitored the two different definitions of the
Poisson ratio, namely, Eq.\ (\ref{eq:PoissonCalc}) using $g_{11}$ in the
denominator, or else using $g_{22}$. Figs.\ \ref{fig:Poisson1} and
\ref{fig:Poisson2} show the values of $\sigma$ as obtained using these two definitions,
for three different conditions on $\kappa$ within the flat phase.
For $\kappa=1.1$, which is near the crumpling
transition, we have found lack of isotropy for almost all simulated
lattice sizes, see Fig.\ \ref{fig:Poisson1}. However, we can safely assume isotropy for
$\kappa=2$ and $L>32$, and for $L>16$ if $\kappa=3.0$, see Figs.\ \ref{fig:Poisson1} and
\ref{fig:Poisson2}. In the following, we will only consider
$\kappa=2,3$, and values of the lattice size for which isotropy holds.

In order to extract the asymptotic value of the Poisson ratio, we will use
the standard scaling form,
\begin{equation}
\sigma(L)=\sigma_\infty+ \frac{A}{L^\omega} ,
\label{eq:PoissonFit}
\end{equation}
where $\omega$ corresponds to the leading correction-to-scaling exponent in
the flat phase. We recall that the analytical prediction is
$\sigma_\infty=-1/3$ \cite{Doussal92}.

We have obtained very good fits for both values of $\kappa=2,3$. Specifically,
for $\kappa=2$ we obtain $\sigma_\infty=-0.30(1)$ and $\omega=1.3(9)$
($\chi^2/\mathrm{d.o.f.}=2.30/2$), while $\kappa=3$ leads to
$\sigma_\infty=-0.31(2)$ and $\omega=0.76(47)$
($\chi^2/\mathrm{d.o.f.}=4.8/3$). In both cases the asymptotic value of the
Poisson ratio is fully compatible with the analytical prediction. We can try
to improve our analysis by fixing $\sigma_\infty=-1/3$. For $\kappa=2.0$ this
leads to $\omega=0.44(6)$ ($\chi^2/\mathrm{d.o.f.}=4.8/3$), while for
$\kappa=3$ we obtain $\omega=0.46(5)$ ($\chi^2/\mathrm{d.o.f.}=4.4/4$).

Finally, we have tried a simultaneous fit of the $\kappa=2$ and 3 data, i.e.,
a fit in which we assume that the values of $\sigma_\infty$ and $c$ in
Eq.\ ({\ref{eq:PoissonFit}) are the same for both $\kappa$, and in which we
  allow for different values of $A$. The result of this joint fit is
  $\sigma_\infty=-0.317(12)$ and $\omega=0.63(20)$
  ($\chi^2/\mathrm{d.o.f.}=8.4/7$). Further, a joint fit in which we fix
  $\sigma_\infty=-1/3$ leads to $\omega=0.46(2)$
  ($\chi^2/\mathrm{d.o.f.}=9.6/8$). Figure \ref{fig:Poisson3} displays the
  linear dependence of the Poisson ratio with $1/\sqrt{L}$.

\begin{figure}[t!]
\centering
\includegraphics[width=0.72\columnwidth, angle=270]{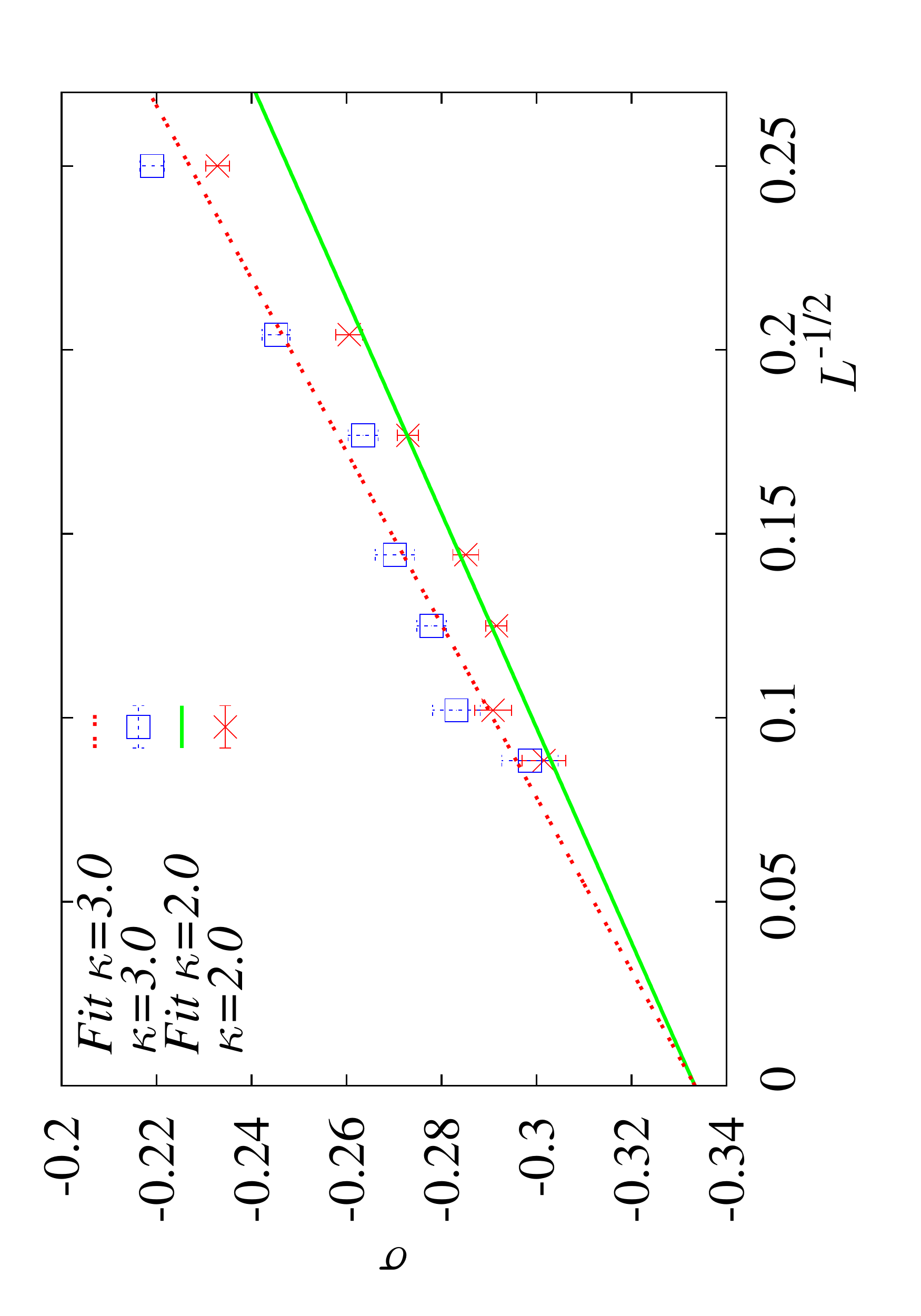}
\caption{(color online) Finite-size effects in the Poisson ratio for
  $\kappa=2$ and 3. We have fixed the scaling correction exponent to 1/2 and
  the asymptotic value to the analytical value, $\sigma_\infty=-1/3$. We have
  used crosses and solid line fit ($\kappa=2$) and squares and dashed line fit
  ($\kappa=2$). }
\label{fig:Poisson3}
\end{figure}

\section{Discussion and conclusions}

We have studied in detail important universal properties associated with the flat
phase of crystalline membranes, specifically the nature of the crumpling transition
and auxetics. With respect to the former, we have found clear signatures for a second
order phase transition. The values we obtain for the critical exponents rule out
completely an interpretation in terms of a strong, or even a weak, first order phase
transition, as had been proposed in the literature recently.

In order to asses the crumpling transition as a continuous one, we have
studied the critical behavior of the specific heat and of the
$\kappa$-derivative of the gyration radius.  In our numerical analysis, we
have employed realistic boundary conditions for the membrane (free boundary
conditions). We have also avoided restricting the computation of our
observables to an inner region far from the boundary, as done elsewhere, see
e.g.\ Ref.\ \cite{Bowick96}.

The values we obtain for the critical exponents [$\alpha=0.55(2)$,
  $\nu=0.73(1)$, $\nu_F=0.74(1)$, and $d_H=2.70(2)$] compare very well with
previous results. For the specific-heat exponent, we can mention
$\alpha=0.5(1)$ \cite{Bowick96}, 0.58(10) \cite{Wheater96}, and 0.44(5)
\cite{Renken90}. For the correlation length exponent, $\nu=0.72(2)$
\cite{Renken90}, 0.68(10) \cite{Wheater96}, and 0.85(14)
\cite{Espriu96}. Regarding the value of the critical coupling, we can quote
$\kappa_c=0.814(2)$ \cite{Wheater96}, to be compared with our result,
$\kappa_c=0.773(1)$. Finally, the value for the Flory exponent or,
equivalently, for the Haussdorf dimension, should be compared with the
analytical results $\nu_F=0.73$, $d_H=2.73$ \cite{Doussal92} and those from
numerical simulations, $\nu_F=0.71(3)$, $d_H=2.77(10)$ \cite{Espriu96}.

Regarding comparison with previous results, we believe that in many cases the
error bars provided for the critical exponents reported in the literature have
been clearly underestimated. We have computed the error bars after a carefully
study of the different integrated correlation times, having simulated among
the largest available lattices. In spite of this, e.g.\ the value of the
critical coupling reported by some previous numerical works is incompatible
with our numerical result, which points out a potential inadequacy of the
corresponding statistical analysis of numerical data. For instance,
$\kappa_c\simeq 0.82$ \cite{Espriu96} and 0.814(2) \cite{Wheater96} are
previous determinations of the critical coupling which are 18 standard
deviations from the value we obtain.

Beyond the nature and critical exponents of the crumpling transition, we have
further studied the value of the Poisson ration in the flat phase as a further
universal property of CM. This seems particularly interesting in view of the
huge scientific and technological importance that auxetic materials are
displaying in recent years \cite{Greaves11}.

In particular, we have simulated three different values of $\kappa$ above the
crumpling transition and in the flat phase, in order to compute the asymptotic
value of the Poisson ratio. We have discarded the smallest $\kappa$ value due to
the huge anisotropies that occur for almost all simulated lattice sizes.
By using the largest two values of $\kappa$, we have found a very precise
infinite-volume extrapolation for the Poisson ratio which is in very good
agreement with the analytical value $\sigma_\infty=-1/3$ computed in
Ref.\ \cite{Doussal92}. In order to reach this conclusion, control of scaling
corrections has been required. We have specifically found numerically that the
scaling corrections behave as $1/\sqrt{L}$. Our extrapolated value for
$\sigma$ is compatible with previous numerical work
\cite{Bowick96,Falcioni97,Bowick01_c} which postulates such a value as
characteristic of a universality class of membranes with fixed connectivity.

In principle, graphene is believed to provide a conspicuous experimental
realization of crystalline membranes. In this context, e.g.\ the model
provided by Eq.\ (\ref{ELandau}) is being intensively employed on a
phenomenological basis as a prototype to describe the statistical-mechanical
properties of this important material system
\cite{Katsnelson12,Katsnelson2013}. Nevertheless, some of the behaviors of
graphene like, for instance, the temperature dependence of the bending
rigidity, seem to remain beyond this type of phenomenological approach
\cite{Katsnelson12,Katsnelson2013}. The fact that the Poisson ratio of
graphene is positive unless, e.g., temperature is sufficiently high
\cite{Zakharchenko2009} or defects are introduced into the crystalline
structure \cite{Grima15}, suggests that predicting realistic values of
$\sigma$ also remains beyond current phenomenological models of graphene
membranes. From the theoretical point of view, it will be interesting to
identify which is the nature of the modifications to be made on generic models
like Eq.\ (\ref{ELandau}) so that they can eventually improve upon this type
of predictions.

\section{Acknowledgments}

This work was partially supported by Ministerio de Econom\'{\i}a y
Competitividad (Spain) through Grants No.\ FIS2012-38866-C05-01 and
No.\ FIS2013-42840-P, by Junta de Extremadura (Spain) through Grant
No.\ GRU10158 (partially funded by FEDER), and by European Union through Grant
No.\ PIRSES-GA-2011-295302. We also made use of the computing facilities of
Extremadura Research Centre for Advanced Technologies (CETA-CIEMAT), funded by
the European Regional Development Fund (ERDF).

\appendix

\section{Some elasticity equations}
\label{app1}
In this appendix we recall some standard elasticity equations and use
them to obtain Eq.\ (\ref{eq:PoissonCalc}), which has been employed in this work to compute
numerically the Poisson ratio.

The starting point is Eq.\ (\ref{eq:elastic}), considered as a Hamiltonian, see
Ref.\ \cite{Lubensky88,Falcioni97}. This equation can be rewritten as
\begin{eqnarray}
\nonumber
{\cal H}_E(u_{\alpha, \beta})&=&\int d^2x \left[ \mu \left( u_{\alpha \beta} -\frac{1}{2} \delta_{\alpha
  \beta}\right)
\left( u^{\alpha \beta} -\frac{1}{2} \delta^{\alpha \beta}\right)\right.\\
&+& \left.\frac{1}{2} K u_\alpha^\alpha u_\beta^\beta \right] \,,
\end{eqnarray}
where $K=\lambda+\mu$ is the compressibility modulus.
From this equation we can compute the stress tensor, $\sigma_{\alpha
  \beta}$ \cite{stress}
\begin{eqnarray}
\nonumber
\sigma_{\alpha \beta}(\mathitbf{x})&=&\langle \frac{\delta {\cal H}_E}{\delta{u^{\alpha \beta}}(\mathitbf{x})}\rangle=
K \delta_{\alpha \beta} \langle u_\gamma^\gamma(\mathitbf{x}) \rangle\\
&+& 2 \mu \left( \langle u_{\alpha
  \beta}(\mathitbf{x}) \rangle-\frac{1}{2} \delta_{\alpha \beta} \langle u_\gamma^\gamma(\mathitbf{x})
\rangle   \right)\,.
\end{eqnarray}
Using that $\sigma_\gamma^\gamma= 2 K \langle u_\gamma^\gamma \rangle$, it is possible to
invert the last equation, obtaining
\begin{equation}
\label{eq:us}
\langle u_{\alpha \beta}(\mathitbf{x}) \rangle =\frac{1}{4 K} \delta_{\alpha \beta} \sigma_\gamma^\gamma(\mathitbf{x})+ \frac{1}{2 \mu} \left(
\sigma_{\alpha \beta}(\mathitbf{x}) -\frac{1}{2} \delta_{\alpha \beta} \sigma_\gamma^\gamma(\mathitbf{x})
\right) \,.
\end{equation}
The Poisson ratio can be written as \cite{Landau75}
\begin{equation}
\label{eq:PoissonDef}
\sigma=-\frac{\langle u_{22} \rangle}{\langle u_{11} \rangle}=\frac{K-\mu}{K+\mu} \,,
\end{equation}
where the directions $x^1$ and $x^2$ on the substrate (the
$\mathitbf{x}$-plane) on which lives the flat
surface are denoted by the indices 1 and 2, respectively.

Using the linear response theorem, we can write~\cite{Falcioni97}
\begin{equation}
\langle u_{\alpha \beta}(\mathitbf{x}) u_{\gamma \delta}(\mathitbf{y}) \rangle_c=
\frac{\delta \langle u_{\alpha \beta}(\mathitbf{x}) \rangle }{\delta \sigma^{\gamma
    \delta}(\mathitbf{y})} \,,
\end{equation}
where $\langle (\cdot \cdot \cdot)\rangle_c$ denotes the connected average as elsewhere in this work.
In addition, taking the derivative of Eq.\ (\ref{eq:us}), one can obtain
\begin{eqnarray}
\nonumber
\frac{\delta \langle u_{\alpha \beta}(\mathitbf{x}) \rangle }{\delta \sigma^{\gamma
    \delta}(\mathitbf{y})}&=&\left( -\frac{K-\mu}{4 \mu K} \delta_{\alpha \beta} \delta_{\gamma
  \delta} \right.\\
&+& \left. \frac{1}{4 \mu} \left(\delta_{\alpha \gamma} \delta_{\beta
    \delta}+ \delta_{\alpha \delta} \delta_{\beta \gamma}  \right)\right) \delta(\mathitbf{x}-\mathitbf{y}) \,.
\end{eqnarray}
Notice that, in a continuous system with a {\em finite} area, $\delta(\mathitbf{0})=1$ \cite{Lubensky2000}. Using the last two equations, we can write
\begin{equation}
\langle u_{11}(\mathitbf{x}) u_{22}(\mathitbf{x}) \rangle_c=
\frac{K-\mu}{ 4 \mu K}\,,
\end{equation}
\begin{equation}
\langle u_{12}(\mathitbf{x}) u_{12}(\mathitbf{x}) \rangle_c= \frac{1}{ 4 \mu} \,,
\end{equation}
\begin{equation}
\langle u_{11}^2(\mathitbf{x}) \rangle_c=\langle u_{22}^2(\mathitbf{x}) \rangle_c=
\frac{K+\mu}{ 4 \mu K} \,.
\end{equation}
Hence, we can finally write Eq.\ (\ref{eq:PoissonDef}) as
\begin{equation}
\sigma=\frac{K-\mu}{K+\mu}=-\frac{\langle u_{11} u_{22} \rangle_c}{\langle
  u_{11}^2 \rangle_c}=-\frac{\langle u_{11} u_{22} \rangle_c}
{\langle u_{22}^2 \rangle_c} \,,
\end{equation}
which, when written in terms of the metric $g_{\alpha \beta}(\mathitbf{x})$, gives us
Eq.\ (\ref{eq:PoissonCalc}) of the main text.

Note that we can compute the tangent vectors, $\partial_i \vec{r}$, as
differences. Here, $i=1,2,3$ run on the three natural directions of the
triangular lattice, which are not orthogonal, while in the definition of
$\sigma$ we assume that the deformations are mutually orthogonal
\cite{Landau75}. Hence, we have defined two orthogonal axes $\mathitbf{x}_1$
and $\mathitbf{x}_2$, taking $\mathitbf{x}_1=\mathitbf{e}_1$ and
$\mathitbf{x}_2=(\mathitbf{e}_2+\mathitbf{e}_3)/\sqrt{3}$, where
$\mathitbf{e}_1$, $\mathitbf{e}_2$ and $\mathitbf{e}_3$ are the three natural
unit vectors on the triangular lattice.  Finally, we compute the induced
metric, $g_{\alpha \beta}$, in this basis.

\section{RG computation of the correction-to-scaling exponent for the flat phase}
\label{omega}
We can obtain a prediction within the RG framework, for the value of the correction-to-scaling exponent for the flat phase. Following Refs.\ \cite{Lubensky88} and \cite{Bowick01}, within an $\epsilon$-expansion the renormalized (dimensionless) elastic constants, $\hat \mu$
and $\hat \lambda$, have the following $\beta$-functions (denoted as $\beta_{\hat \mu}$ and
$\beta_{\hat \lambda}$):
\begin{eqnarray}
\beta_{\hat \lambda}&=& \frac{ d {\hat \lambda}}{ d \log l}=
-\epsilon {\hat \mu} + \frac{{\hat \mu}^2}{8 \pi^2} \left(\frac{1}{3} d_c +20 A
  \right)\,, \label{eq:bl} \\
\beta_{\hat \mu}&=& \frac{ d {\hat \mu}}{ d \log l}=
-\epsilon {\hat \lambda}+ \frac{1}{8 \pi^2}\left[\frac{1}{3} d_c {\hat \mu}^2 \right. \nonumber \\ &+& \left. 2(d_c+10 A) {\hat \lambda} {\hat \mu}+2 d_c {\hat \lambda}^2 \right]\, . \label{eq:bm}
\end{eqnarray}
Here, $A=({\hat \mu}+{\hat \lambda})/(2 {\hat \mu}+{\hat \lambda})$.
In our present case, $\epsilon=2$ and $d_c=1$ \cite{Bowick01}. Eqs.\ (\ref{eq:bl})-(\ref{eq:bm}) present four fixed points. Out of these, the physical behavior is controlled by
the one characterized by ${\hat \lambda}=-4 \epsilon/(24+d_c)$ and ${\hat \mu}=12 \epsilon/(24+d_c)$.

In order to compute the correction-to-scaling exponent, we need to obtain the
eigenvalues, denoted by $\lambda_1$ and $\lambda_2$, of these two couplings via the diagonalization of the Jacobian matrix of the $\beta$-functions \cite{Amit2005}. Note, both eigenvalues are expected to be negative, as the system lies on the critical surface. Indeed, the result is
\begin{eqnarray}
\lambda_1&=&\frac{(65-125 \pi^2) \epsilon}{125 \pi^2}\,,\\
\lambda_2&=&\frac{125 (1- \pi^2) \epsilon}{125 \pi^2}\,.
\end{eqnarray}
Setting $\epsilon=2$, we obtain $\lambda_1\simeq-1.90$ and
$\lambda_2\simeq-1.80$. Hence, $\omega\simeq-1.80$ at this perturbative
order. 

Finally we can compute the value of the Poisson ratio, by using that, at the
physical fixed point, $\hat \lambda/ \hat \mu=-1/3$, thus $\sigma=\hat \lambda/(\hat \lambda+2 \hat \mu)=-1/5$. Notice that both values for $\omega$ and $\sigma$ are far from the corresponding MC results. We can conclude that, due to the large value of $\epsilon$, it is very difficult to
extract reliable predictions from the RG at this perturbative order.

\section{Implementation of the code in GPU}
\label{gpu}

As widely acknowledged \cite{Wei2012}, pretty good gain factors can be
obtained by using Graphic Processing Units (GPU). In our case, the gain was
almost guaranteed, given that we use mostly floating point variables that can
be updated in parallel following some kind of checkerboard algorithm. The gain
factor depends strongly on the system size, but we obtain at least a factor of
5x.

The Compute Unified Device Architecture (CUDA) implementation of the
Metropolis algorithm stores a single copy of the surface in the GPU memory. By
operating only in the GPU memory, we reduce memory controller load and reduce
the computation time required per iteration. The surface update kernel is
executed sequentially over a quarter of the surface nodes and returns to CPU
mode once all nodes have been processed, see Fig.~\ref{fig:GPU}.

The structure of the algorithm requires performing an atomic operation over a
single variable to sum the total number of updated node positions. This is the
hardest bottleneck in the algorithm.

Due to the parallel update structure, we need to generate a random seed for
each node in the surface, as opposed to the CPU sequential version, in which a
single seed is enough. The number of blocks and threads per block in the CUDA
simulation is given by the surface size and the maximum number of GPU
registers assigned to each thread. The limit in the number of registers that a
kernel can use has a great impact on the overall performance.


\begin{thebibliography}{99}
\bibitem{metric} Notice that, using the flat membrane parameterization (Monge parameterization),
$\vec{r}=(\mathitbf{y}, h(\mathitbf{y}))$, the metric tensor can be evaluated as
$$
g_{\alpha \beta}=\delta_{\alpha \beta} +\frac{\partial^2 h}{\partial y^\alpha
  \partial y^\beta} .
$$
Hence, to order $O(h)$, one can freely raise and lower incides. Moreover, $d^2 x= d^2 s$.
\bibitem{extrinsic} The extrinsic curvature matrix is defined in the usual way \cite{David04}: The
derivative of a tangent vector can be expressed in the (local) basis of the
membrane tangent vectors and normal vector (denoted as $\vec{n}$) as
$$
\partial_\alpha \vec{t}_{\beta}= \Gamma^\gamma_{\alpha \beta} \vec{t}_\gamma +
K_{\alpha \beta} \vec{n},
$$
where $\Gamma^\gamma_{\alpha \beta}$ are the Christoffel symbols (affine
connection).
\end{thebibliography}
\end{document}